\documentclass[conference]{IEEEtran}
\IEEEoverridecommandlockouts
\usepackage{cite}
\usepackage{amsmath,amssymb,amsfonts}
\usepackage{graphicx}
\usepackage{textcomp}
\usepackage{xcolor}
\usepackage{colortbl}
\usepackage{array}
\usepackage{listings}
\usepackage{xcolor}
\usepackage{tabularx}
\usepackage{enumitem}

\usepackage{algorithm}
\usepackage{algpseudocode}
\usepackage{svg}

\makeatletter

\def\ps@IEEEtitlepagestyle{%
  \def\@oddfoot{\mycopyrightnotice}%
  \def\@evenfoot{}%
}
\def\mycopyrightnotice{%
  {\footnotesize 979-8-3503-8532-8/24/\$31.00 ©2024 IEEE\hfill}
  \gdef\mycopyrightnotice{}
}

\def\BibTeX{{\rm B\kern-.05em{\sc i\kern-.025em b}\kern-.08em
    T\kern-.1667em\lower.7ex\hbox{E}\kern-.125emX}}
\begin{document}

\title{Optimizing Cloud-native Services with SAGA: A Service Affinity Graph-based Approach}

\author{
    \IEEEauthorblockN{Hai Dinh-Tuan}
    \IEEEauthorblockA{
        \textit{Service-centric Networking} \\
        \textit{Technische Universität Berlin} \\
        Berlin, Germany \\
        hai.dinhtuan@tu-berlin.de\\
    }
    \and
    \IEEEauthorblockN{Franz Florian Six}
    \IEEEauthorblockA{
        \textit{Technische Universität Berlin} \\
        Berlin, Germany \\
        six.florian00@gmail.com
    }
}

\maketitle

\begin{abstract}
Modern software architectures are characterized by their cloud-native, modular, and microservice-based designs. While these systems are known for their efficiency, they also face complex challenges in service optimization, especially in maintaining end-to-end quality of service across dynamically distributed services. This paper introduces a novel approach using the concept of \textit{Service Affinity} to address this challenge. The proposed method, termed \textit{Service Affinity Graph-based Approach}, employs a graph-based model to model the interactions among microservices. It formulates the service placement as a minimum-weight k-cut problem and utilizes an approximation algorithm for service clustering. This approach is realized through a conceptual framework that takes into account a wide range of optimization objectives, ranging from enhancing application performance and enforcing data privacy to optimizing operational costs. In addition to presenting the SAGA framework in details, this paper conducts an in-depth empirical evaluation using a prototype deployed on a Kubernetes cluster. The results demonstrate a mean latency improvement of 23.40\%, validating the effectiveness of our approach. Finally, the paper comprehensively discusses various aspects of the proposed methods, including their implications, challenges, and benefits, providing a thorough analysis of the approach's impact.

\end{abstract}

\begin{IEEEkeywords}
microservices, cloud-native, kubernetes, service optimization
\end{IEEEkeywords}

\IEEEpubidadjcol

\section{Introduction}

In the evolving landscape of software engineering, the notion of scalability has assumed a central role, enabling cloud-based services to serve millions of users across geographical regions. This trend is driven by three key aspects of modern software architectures: cloud-native design, modularity, and a microservice-based approach. Cloud-native applications utilize the cloud's abstracted computing resources for flexible deployment across various hosts. Modularity now extends beyond design to runtime, separating processes and network communications for individual modules, diverging from traditional shared-resource models. Finally, the shift towards a microservice-based architecture reflects a preference for smaller, independent modules, promoting efficient deployment and supporting large team management \cite{dinh2019maia}. However, alongside these advantages, the decentralization in modern software architectures also brings forth a set of new challenges. These challenges include, but are not limited to, resource allocation, cost management, dynamic scaling under fluctuating workloads, and the assurance of end-to-end quality of service. 

Existing cloud orchestration tools mainly offer service placement features focused on resource management, involving configurations of computing power, memory, and storage. Although these tools allow adjustments in the allocation of services to designated nodes, they fall short in addressing the dynamic aspects of services, including their interdependencies, data exchange volumes, and functional relationships.

This research seeks to address this challenge by proposing a novel approach centered around the concept of \textit{Service Affinity (SA)}. Tailored for cloud-native environments, this method differs itself from traditional resource-centric models by integrating application-specific metrics such as data exchange volumes, the nature and quantity of data, privacy requirements, and functional dependencies among microservices. It aims to improve service management by aligning resource allocation with application-specific requirements, thereby enhancing performance and efficiency. The proposed \textit{Service Affinity Graph-based Approach (SAGA)} gathers data from various cloud-native application components, including messaging infrastructure and cloud orchestrators. Utilizing this metadata, SAGA applies graph theory to construct a model for the microservice cluster. This graph-based representation reformulates the service placement as a minimum-weight k-cuts problem. To address this, an approximation algorithm is employed, ensuring a solution that is near-optimal while also being computationally practical within a feasible runtime.

The contributions of this paper are therefore two-fold:
\begin{itemize}
    \item \textit{Development of the SAGA conceptual framework:} At the core of this framework lies the SA concept, designed to streamline the optimization of cloud-native services. This framework aligns optimization objectives with diverse requirements from multiple stakeholders, including software developers and cloud operators, enabling a collaborative approach to service optimization.

    \item \textit{Implementation and Evaluation in a real environment:} The research goes beyond theoretical development by implementing and evaluating the concept through a prototype in Kubernetes, a de facto standard service orchestration platform. This practical evaluation proves the proposed concept feasibility and provides insights into its practical applications, highlighting benefits and challenges.
\end{itemize}

The remainder of this paper is organized as follows: Section II commences with a review of related work in the domain of service optimization in the cloud context. The algorithms for SA calculation as well as the k-way extension of  Kernighan-Lin algorithm are thoroughly explained in Section III. Section IV is devoted to presenting our prototype implementation of SAGA in Kubernetes with both quantitative and qualitative evaluation results. Lastly, Section V not only concludes our findings but also paves the way for future research directions.

\section{Related work}

Service placement in the domains of microservices and cloud/edge computing represents a complex challenge. It involves optimizing hosting locations to achieve a range of objectives, including but not limited to, minimizing latency, balancing workloads, enhancing cost efficiency, and ensuring high availability. The complexity of this challenge is further exacerbated by the distributed nature of these systems and involves tackling NP-hard combinatorial optimization problems. Traditional deterministic algorithms, while effective in certain contexts, often fall short in terms of scalability when applied to these complex problems. Consequently, there's a shift towards dynamic, adaptive methodologies. This section reviews some of those works, categorized by their approaches. 

One example of heuristic methods is the \textit{Iterative Greedy-and-Search-based algorithm (IGS)} proposed by \cite{fan2021cloud}, designed for optimizing resource allocation and pricing in cloud/edge computing for mobile blockchain applications. The work addresses the Stackelberg game between Cloud/Edge Service Providers (CESP) as the leader and users as followers for resource management, proving the existence of equilibrium and presenting a mixed integer programming formulation for the problem. The algorithm addresses two key sub-problems: resource allocation under specified pricing, and resource pricing based on the allocation scheme. Utilizing a greedy-and-search approach for the former and golden section search for the latter, the IGS algorithm effectively enhances revenue for both CESP and users. The simulation results underscore the algorithm's efficiency in increasing revenues and reducing task execution delay, particularly as the user revenue parameter $\alpha_i$ and server computing power increase.

Compared to heuristic methods, exact methods for solving service allocation problems typically involve approaches that guarantee finding the optimal solution, albeit often at the cost of high computational complexity. Liu and Fan \cite{liu2018resource} address resource allocation in a multi-cloudlet environment using a two-stage  strategy based on Mixed Integer Linear Programming (MILP). The first stage involves selecting the optimal cloudlet by considering latency and rewards, while the second stage allocates resources within the selected cloudlet to optimize system reward and resource usage. Their MILP formulations focus on maximizing mean reward and minimizing latency, with simulations showing enhanced system performance. 

Metaheuristic methods for resource allocation are a class of optimization algorithms used to solve complex resource allocation problems in a reasonable time frame. These methods are especially valuable for large-scale or dynamic environments where finding an exact solution is computationally infeasible. The authors in \cite{huang2020new} present an automatic smart service migration framework using the ant colony optimization (ACO) algorithm on cloud-oriented e-commerce. This is a three-step process: (1) the initial population phase, where algorithm parameters are initialized and to-be-migrated services are randomly allocated to servers; (2) the solution creation phase, where the ACO algorithm is applied to generate a better solution (low cost and energy consumption); and (3) the pheromones update phase, where the pheromone concentration of the edges that lead to better solutions is increased and the pheromone concentration of the edges that lead to worse solutions is decreased.

There are works that employ mathematical methods, like the \textit{bi-objective optimization model} presented by Bento et al. \cite{bento2022bi}. This model aims to balance service availability and cost, maximizing the former while minimizing the latter, and considers infrastructure capacity and service-level objectives. Using mathematical formulations and a Pareto front approach, it identifies optimal solutions for system configuration. A case study demonstrates its effectiveness in enhancing service availability and cost efficiency, indicating a shift towards proactive, global optimization of cloud services, as opposed to traditional reactive, local methods.

Other studies adapt the graph theory, for example, Devi and Murugaboopathi\cite{devi2019efficient} introduce the \textit{Cloud Load Balancer (CLB)} algorithm , which includes \textit{Cluster Data Center Clustering (CDCC)} and \textit{Client Cluster Assignment (CCA)}. CLB is compared with graph-based k-means and k-spanning tree algorithms, focusing on efficient clustering of distributed cloud data centers. Notably, CLB's CDCC algorithm selects initial cluster centers based on node degree, leading to reduced clustering times. The CCA method ensures a more uniform distribution of clients across clusters. Overall, CLB shows significant improvements in service delivery performance and load balancing in multi-cloud environments.

There are also other works focusing on actual techniques for service migration within cloud-native environments, which not only leads to more efficient use of resources~\cite{guo2022performability} but also helps overcome challenges like network congestion or server downtime, ensuring uninterrupted service availability. For example, the \textit{Message-based Stateful Microservice Migration (MS2M)} \cite{dinh2022ms2m}, utilizes the messaging infrastructure, integral to many microservice applications, to manage the service's state during migration. This approach reduces downtime while ensure seamless service availability during migration. Another worth mentioning work is \textit{Sledge} \cite{xu2020sledge}, which introduces a \textit{Lightweight Container Registry (LCR)} mechanism for end-to-end image migration and employs \textit{Dynamic Context Loading (DCL)} for management context migration. Sledge's approach significantly reduces migration downtime, total migration time, and image migration time compared to existing solutions. 

\section{Concept and Design}

In this section, we introduce the concept and design of our modified k-way Kernighan-Lin Algorithm aimed at optimizing microservice deployment in a distributed edge-cloud environment. The core idea revolves around effectively partitioning a graph into $k$ balanced subsets, where each subset represents a cluster of microservices with high internal affinity and minimal inter-cluster interactions. The detailed algorithmic steps are outlined below, followed by a comprehensive explanation of the affinity types and their relevance in determining the optimal clustering configuration.

\begin{algorithm}
\caption{k-Way Kernighan-Lin Algorithm}\label{kla:cap}
\begin{algorithmic}[1]
\Procedure{Kernighan-Lin-k}{$G(V, E)$, $k$}
    \State Initialize a list $S$ containing all vertices
    \While{length of $S$ $< k$}
        \State Identify the largest subset $L$ in $S$
        \State Partition vertices in $L$ into balanced sets $A$ and $B$
        \Repeat
            \State Compute $D$ values for all $a \in A$ and $b \in B$
            \State Let $gv$, $av$, and $bv$ be empty lists
            \For{$n := 1$ \textbf{to} $|L| / 2$}
                \State Find $a \in A$ and $b \in B$ to maximize $g = D[a] + D[b] - 2 \times c(a, b)$
                \State Remove $a$ and $b$ from further consideration in this pass
                \State Add $g$ to $gv$, $a$ to $av$, and $b$ to $bv$
                \State Update $D$ values for the elements of $A = A \setminus \{a\}$ and $B = B \setminus \{b\}$
            \EndFor
            \State Find $t$ which maximizes $g_{\text{max}}$, the sum of $gv[1], \ldots, gv[t]$
            \If{$g_{\text{max}} > 0$}
                \State Exchange $av[1], av[2], \ldots, av[t]$ with $bv[1], bv[2], \ldots, bv[t]$
            \EndIf
        \Until{$g_{\text{max}} \leq 0$}
        \State Replace $L$ in $S$ with two new subsets $A$ and $B$
    \EndWhile
    \State \Return $S$ \Comment{List of k subsets}
\EndProcedure
\end{algorithmic}
\end{algorithm}

Let \( G(V, E) \) be an undirected graph, where each vertex \( v \in V \) represents a microservice instance within an edge-cloud environment, and each edge \( e \in E \) represents a connection between two services, existing solely when data is exchanged between them. . Let \( k \) denote the number of hosts in the cluster. For any pair of nodes \( u, v \in V \), the affinity value \( a_{u,v}(\Delta t) \) during the time window \( \Delta t \) is calculated as:

\begin{equation}
a_{u,v}(\Delta t) = \sum_{i \in \{d, p, c, f, o\}} w_i \cdot a_{i}(u, v, \Delta t)
\label{eq:affinity_value}
\end{equation}

where:
\begin{itemize}
    \item \( a_{u,v}(\Delta t) \) represents the overall affinity value between \( u \) and \( v \) during the time window \( \Delta t \).
    \item \( a_{i}(u, v, \Delta t) \) represents the specific affinity value between \( u \) and \( v \) for type \( i \), with \( i \) being data (\( d \)), privacy (\( p \)), coupling (\( c \)), functional (\( f \)), and operational (\( o \)), during the time window \( \Delta t \).
    \item \( w_i \) is the weight assigned to each affinity type.
\end{itemize}

The weights assigned to different affinity types are adjustable according to the specific environment, allowing fine grain control over the overall affinity calculation. Further details of each affinity type are elaborated in the subsequent sections.

\subsection{Data Affinity (d)}
Data Affinity measures the extent of data exchange between two microservices. High data affinity indicates frequent sharing or transfer of significant data amounts. To calculate the data affinity (\( d_{u,v} \)) between two microservices \( u \) and \( v \), the formula is as follows:

\[d_{u,v}(\Delta t) = \frac{b_{u,v}(\Delta t)}{B(\Delta t)}\]

where:
\begin{itemize}
    \item \( d_{u,v}(\Delta t) \) is the Data Affinity between services \( u \) and \( v \) during the time window \( \Delta t \).
    \item \( b_{u,v}(\Delta t) \) represents the total bytes of data exchanged between microservices \( u \) and \( v \) during the time window \( \Delta t \),
    \item \( B(\Delta t) \) denotes the total bytes of data exchanged across the entire application during the time window \( \Delta t \).
\end{itemize}

\subsection{Privacy Affinity (p)}
Privacy Affinity enables precise microservice placement control in alignment with data privacy and security needs. By grouping microservices with similar privacy profiles, this allows users to effectively implement their privacy and security policies, ensuring sensitive data is confined to a specified subset of hosts within the edge-cloud environment.

\[
p_{u,v} = 
\begin{cases} 
    1 & \text{if privacy tag of } x = \text{privacy tag of } y, \\
    0 & \text{otherwise.}
\end{cases}
\]

\subsection{Coupling Affinity (c)}
Coupling Affinity reflects the degree of dependency between microservices, indicating how closely one service is tied to another. Let \( c_{u,v}(\Delta t) \) represent the Coupling Affinity between microservices \( u \) and \( v \), calculated over the specific time window \( \Delta t \) as the normalized measure of the number of messages exchanged between them:

\[c_{u,v}(\Delta t) = \frac{m_{u,v}(\Delta t)}{M(\Delta t)}\]

where:
\begin{itemize}
    \item \( c_{u,v}(\Delta t) \) is the Coupling Affinity between services \( u \) and \( v \) calculated over time window (\( \Delta t \)).
    \item \( m_{u,v}(\Delta t) \) represents the number of messages exchanged between \( u \) and \( v \) within the time window (\( \Delta t \)).
    \item \( M(\Delta t) \) denotes the total number of messages exchanged across the entire system within (\( \Delta t \)).
\end{itemize}

\subsection{Functional Affinity (f)}
Functional Affinity measures the similarity in roles or purposes of microservices. Services with high functional affinity have similar functionalities or contribute to the same business process, influencing development and maintenance efficiency. This should be calculated based on the software architecture. The Functional Affinity (\( f_{x,y} \)) is calculated using a simple binary approach:

\[
f_{x,y} = 
\begin{cases} 
    1 & \text{if function tag of } u = \text{function tag of } v, \\
    0 & \text{otherwise.}
\end{cases}
\]

\subsection{Operational Affinity (o)}
Operational Affinity looks at operational aspects like deployment patterns, maintenance requirements, and monitoring needs. Microservices with similar operational profiles can be grouped for efficient management.

\[
o_{u,v} = 
\begin{cases} 
    1 & \text{if operational tag of } u = \text{operational tag of } v, \\
    0 & \text{otherwise.}
\end{cases}
\]

\subsection{Weighted-graph representation of the microservices cluster}

The normalized edge weight \( w_{u,v}(\Delta t) \) for any pair of nodes \( u, v \in V \) is calculated by normalizing the affinity value \( a_{u,v}(\Delta t) \) based on the minimum and maximum affinity values observed across all edges in \( E \) during the time window \( \Delta t \):

\begin{equation}
w_{u,v}(\Delta t) = \frac{a_{u,v}(\Delta t) - \min_{(u,v) \in E} a_{u,v}(\Delta t)}{\max_{(u,v) \in E} a_{u,v}(\Delta t) - \min_{(u,v) \in E} a_{u,v}(\Delta t)}
\label{eq:normalized_edge_weight}
\end{equation}

Formula~(\ref{eq:affinity_value}) and (\ref{eq:normalized_edge_weight}) enable the construction of an undirected, weighted graph \( G(V, E) \), representing the microservice cluster. The objective is to cluster the vertices into $k$ distinct subsets, minimizing the total edge weight between them, which forms a minimum-weight k-cut problem. This NP-hard problem in computer science can be solved ussing several algorithmic strategies, depending on the specific requirments and graph size. For small graphs, exact algorithms like integer linear programming are used, while larger graphs utilize approximation methods like the Greedy and Kernighan–Lin algorithms, or spectral partitioning. For very large graphs, heuristic methods like simulated annealing and genetic algorithms, along with randomized algorithms, are employed for near-optimal solutions.

In our specific case, we have chosen to utilize the Kernighan–Lin Algorithm for its effective balance between solution quality and computational resource efficiency. Originally designed for two-way clustering, we adapted this algorithm to accommodate k-way clustering, enabling the generation of $k$ distinct clusters of nodes as a result. The modified pseudocode is presented in Algorithm~\ref{kla:cap}.

\section{Implementation}

\begin{figure}
	\centering
        \includegraphics[width=\columnwidth]{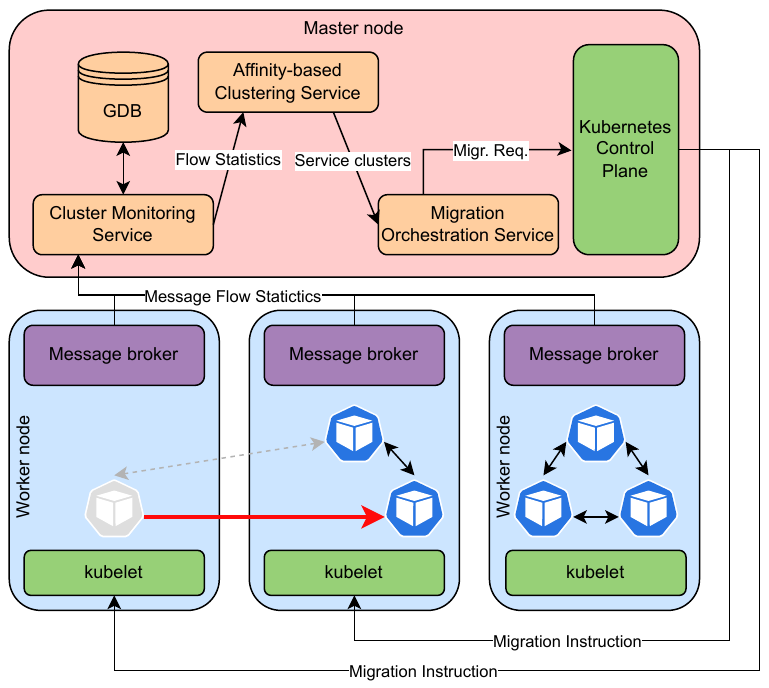}
	\caption{Overview of SAGA Framework prototype integrated within Kubernetes cluster.}
	\label{fig:overview}
\end{figure}

This section presents a prototype implementation of the proposed SAGA framework. The core of the prototype encompasses three services as below: 

\begin{enumerate}
    \item \textit{Cluster Monitoring Service (CMS):} This service is responsible for continuously collecting application data from cloud infrastructures, including communication networks and cloud orchestrators. It primarily tracks and analyzes message flows among microservices, capturing metadata such as sender, receiver, message count, size, and frequency. Additionally, CMS collects other operational metrics through cloud orchestrator APIs.

    \item \textit{Affinity-based Clustering Service (ACS):} ACS is tasked with grouping microservices based on affinity data provided by CMS, focusing on the time window \( \Delta t \). It recalculates and updates the graph representation \( G \) by computing microservice affinity values \( a \) for each pair. This computation employs an extended k-way Kernighan–Lin Algorithm to create \( k \) distinct subsets of microservices, each optimized to minimize total inter-subsets affinity.

    \item \textit{Migration Orchestration Service (MOS):} MOS's role is to adjust service placement in accordance with the clusters determined by ACS. MOS oversees the entire migration process, ensuring the integrity and availability of services. Upon completion of microservice reallocation, it signals CMS to reset its database and start collecting new metric data for the subsequent time window.
\end{enumerate}

While the CMS runs continuously, both ACS and MOS can operate on a set schedule or be triggered on demand, such as in response to increases in system load. Figure~\ref{fig:overview} depicts the high level view of those three services in Kubernetes cluster.

\section{Evaluation and Discussion}

\subsection{Algorithm performance evaluation}

The time complexity for each execution of the Kernighan-Lin algorithm is \( O(n^2 \log n) \). However, the overall complexity for k-way partitioning depends the number of iterations and the sizes of the subsets being partitioned in each iteration. As outlined in Algorithm~\ref{kla:cap}, the algorithm is executed \( \lceil \log_2 k \rceil \) times. This frequency is due to the algorithm's design, where each iteration divides one subset into two, continuing this process until \( k \) subsets are obtained. Therefore, the total estimated complexity for k-way partitioning is \( \lceil \log_2 k \rceil \times O(n^2 \log n) \).

\subsection{Algorithm runtime analysis}

As the theoretical time complexity analysis suggests, the runtime performance of the algorithm depends on the number of microservices \( n \) and the number of nodes \( k \). Therefore, we set up an evaluation where the number of microservices varies from 100 to 500, and the number of nodes ranges from 10 to 50.

\begin{figure}
    \centering
    \includegraphics[width=\columnwidth]{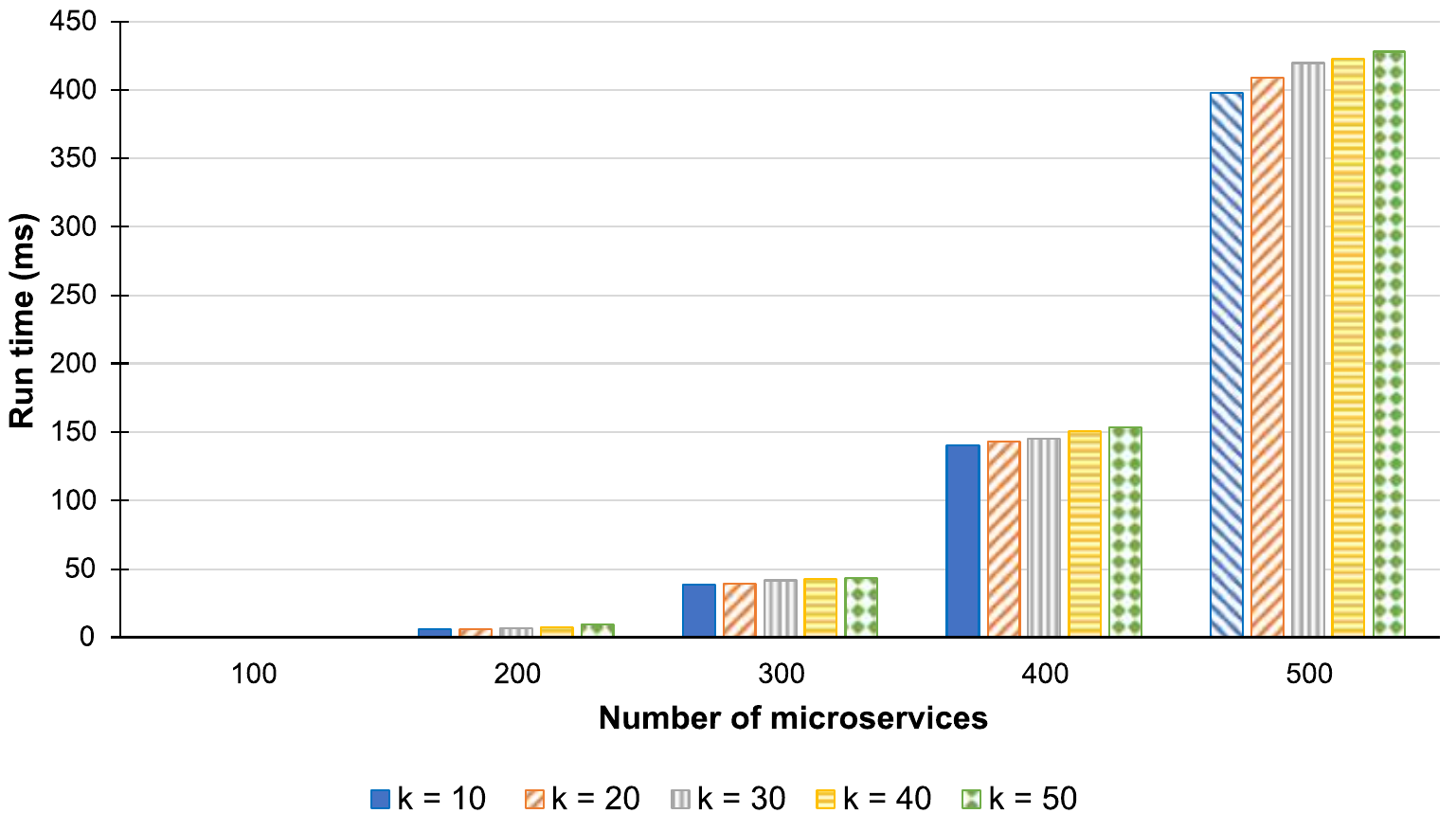}
    \caption{Extended k-way Kernighan-Lin algorithm performance analysis.}
    \label{fig:runtime_analysis}
\end{figure}

The runtime performance analysis of the k-way Kernighan-Lin algorithm, as depicted in Figure~\ref{fig:runtime_analysis}, demonstrates a significant reliance on the number of microservices, complemented by a lesser impact from the cluster count  (\( k \)). For instance, empirical data indicates that in a smaller application with 100 microservices, the runtime ranges from 343.32ms (\( k=10 \)) to 474.83ms (\( k=50 \)). In contrast, for a larger application with 500 microservices, the runtime spans from 398.05s (\( k=10 \)) to 428.00s (\( k=50 \)), highlighting a relatively minor effect of $k$ on algorithm runtime. This can be explained by each run of the algorithm possibly being on a smaller subset of the graph (especially in later runs), meaning the actual computational workload might be less than the time complexity approximation of $\lceil \log_2 k \rceil \times O(n^2 \log n)$ calculated above. Overall, this suggest that the scale of microservices architecture can be a limiting factor for the scalability of the chosen algorithm.

\subsection{Prototype evaluation}

\begin{figure}
	\centering
        \includegraphics[width=\columnwidth]{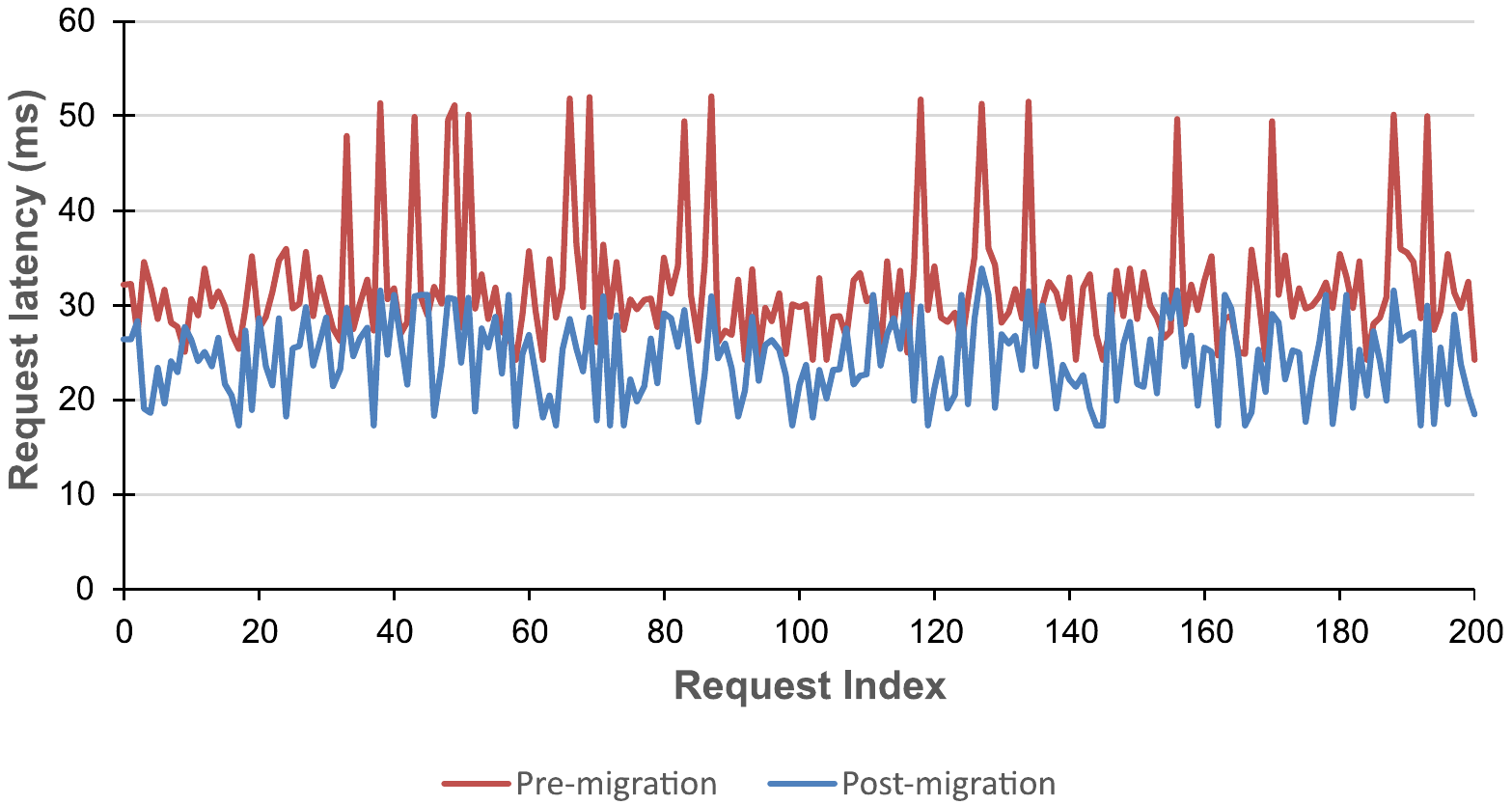}
	\caption{Latency performance of 5 service pairs.}
	\label{fig:latency}
\end{figure}

\begin{figure}
	\centering
    \includegraphics[width=0.6\columnwidth]{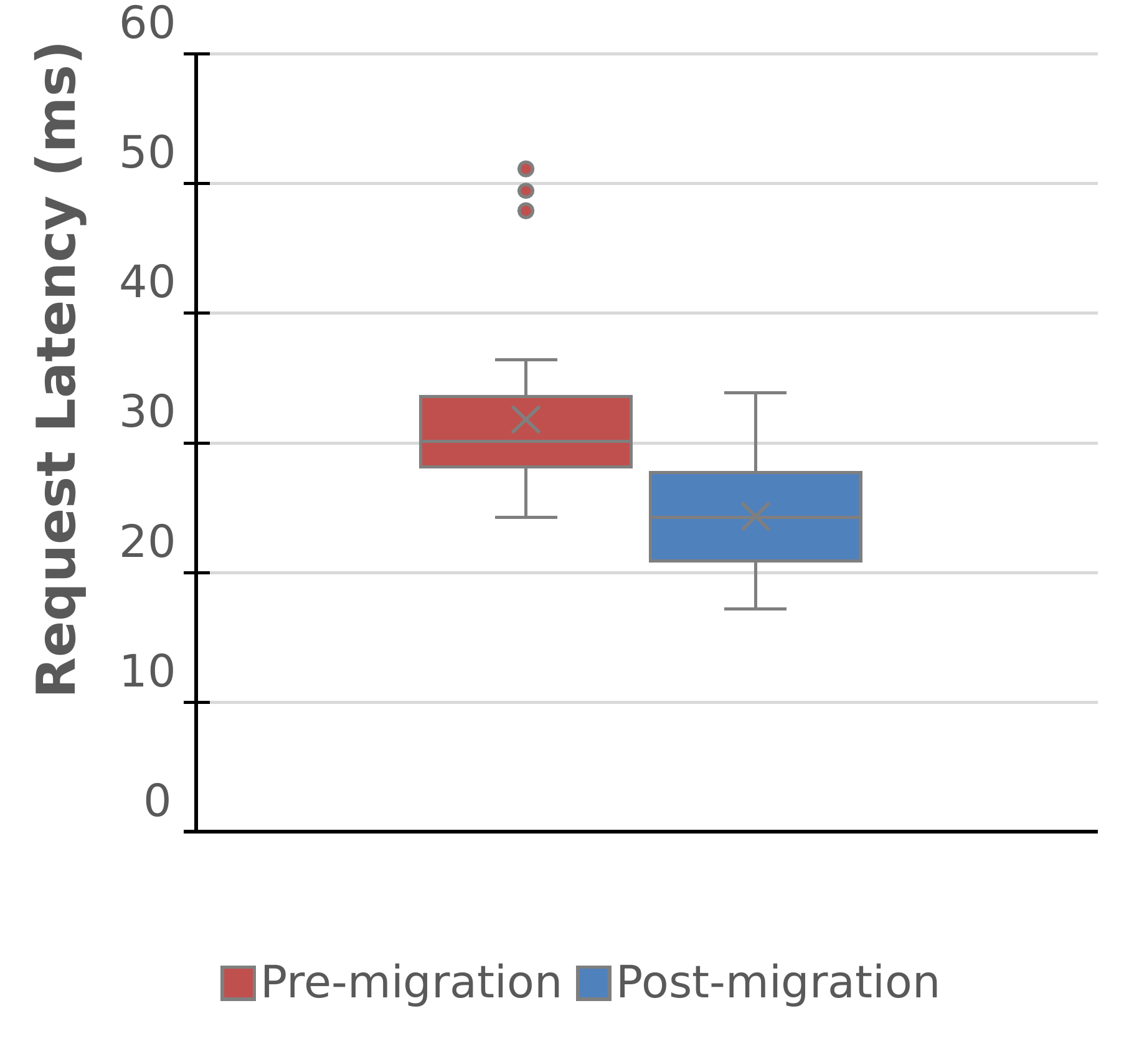}
	\caption{Box plots of the latency performance before and after migration.}
	\label{fig:latency2}
\end{figure}

Our framework was evaluated using Microsoft's open-source .NET Core online-shop application\cite{eshop}, demonstrating basic microservice architecture and employing RabbitMQ for messaging\cite{rabbitmq}. It was deployed on a Kubernetes cluster in the Google Kubernetes Engine\cite{kubernetes}, comprising one master and three worker nodes across different zones for enhanced availability. Each host machine has 8 vCPUs and 32GB of memory, provided by Google Cloud Platform\cite{gcp}. The SAGA framework's components were developed using Java and the Spring Framework\cite{spring}.

In our evaluation, we focused on assessing service latency for tasks like ordering, payment, and item retrieval. Using a dataset of 200 requests, we analyzed latency—defined as the time from request issuance to response receipt—both before and after system migration. To accurately calculate latency performance, we employed distributed tracing techniques\cite{parker2020distributed}, allowing for detailed tracking and analysis of request handling within the system. The results are summarized in Figure~\ref{fig:latency}.

Before migration, the data showed a broaded range of latency values, with minimum and maximum latencies at 24.81 ms and 52.08 ms, respectively. The mean latency was 31.80 ms, with a standard deviation of 6.50 ms, indicating a moderate spread. The interquartile range was from 28.20 ms to 33.54 ms, suggesting some instances of higher latency were noted.

After migration, there was a significant improvement in latency performance. The highest recorded latency was significantly lower at 33.91 ms, and the mean latency reduced to 24.36 ms with a smaller standard deviation of 4.30 ms. The interquartile range also narrowed to 20.98 ms to 27.67 ms. 

The box plot visualization, as seen in Figure~\ref{fig:latency2}, highlights the changes in latency. Pre-migration data showed a wider box and longer whiskers, indicating greater variability in latency. Post-migration, the box plot became more compact with shorter whiskers, suggesting a closer clustering of latency values around a reduced mean.

Overall, the analysis indicates a notable performance improvement post-migration. As depicted in Figure~\ref{fig:latency}, there was a 23.40\% reduction in mean latency, signifying a marked enhancement in latency performance. Additionally, the improvement extends to the variance in request processing latency, demonstrating not just a decrease in average latency but also greater consistency in handling requests.

\section*{Conclusion}

In conclusion, the transition to cloud-native, modular microservices in software architecture demands innovative tools and techniques to maintain service quality. This paper presents the SAGA framework, integrating application-specific metrics with resource-centric models for service optimization, grounded in the concept of service affinity. Utilizing this concept, SAGA constructs a graph-based representation of the microservices cluster and reformulates the service placement as a minimum-weight k-cut problem. The framework extends the Kernighan-Lin algorithm for k-way graph partitioning, effectively addressing this challenge. Empirical evaluation results demonstrate a strong impact of microservices count on the algorithm runtime, while the number of nodes $k$ appears to have a minimal impact, indicating potential scalability limitations of the selected algorithm on large and complex applications.

The implementation of SAGA in a Kubernetes cluster comprises three integral services: Cluster Monitoring Service (CMS) for continuous data collection and analysis of microservice interactions; Affinity-based Clustering Service (ACS) for clustering microservices based on CMS-identified affinities; and Migration Orchestration Service (MOS) for efficient service placement and deployment in coordination with cloud orchestrators, ensuring service integrity and availability.

Initial evaluations of the SAGA prototype have shown a significant enhancement in microservice performance, with a 23.40\% improvement in average request latency, highlighting its potential for optimizing microservice operations through intelligent service placement. SAGA's use of metadata about application behavior, offers a more comprehensive approach to service deployment. This integration enables software developers to influence deployment strategies effectively, aligning the system with specific operational needs and performance goals. By proactively managing resources based on both operational metrics and behavioral insights, developers can anticipate and address potential performance bottlenecks before they affect the user experience. This dynamic adaptability not only improves efficiency but also enhances the agility of service management in cloud environments.

Looking forward, there is considerable scope for expanding the capabilities of the SAGA framework. Among others, improving the efficiency of the algorithms used to solve the minimum-weight k-cut problem is an immediate area of focus. The current implementation of the k-way Kernighan-Lin algorithm shows limited scalability, suggesting a need for more robust and scalable solutions. Its approach of equalizing the number of services in each cluster may hinder the discovery of optimal solutions. Enhancements could include exploring alternative algorithms that offer greater efficiency and scalability.

Another potential focus area is extending the SAGA framework to support multi-domain scenarios, allowing different stakeholders to manage distinct edges and clouds, more accurately reflecting real-world conditions. Additionally, while the SAGA framework effectively identifies subsets of microservices, it currently does not take into account the capabilities of physical nodes, which are crucial for precisely allocating these subsets to appropriate physical nodes.

\bibliographystyle{IEEEtran}
\bibliography{bibliography}
\end{document}